\documentclass[notitlepage,12pt,onecolumn,tightenlines,a4paper,eqsecnum,longbibliography]{revtex4-1}

\usepackage{amsfonts}
\usepackage{amsmath}
\usepackage{amssymb}
\usepackage{bm}
\usepackage{dcolumn}
\usepackage{epsfig}
\usepackage{graphicx}
\usepackage{graphics}
\usepackage[latin1]{inputenc}
\usepackage{latexsym}
\usepackage{rotating}
\usepackage[colorlinks=true]{hyperref}
\usepackage[all]{hypcap} 
\usepackage[margin=1.in]{geometry}
\usepackage{wrapfig}

\usepackage[usenames]{color}

\newcommand{\GR}{{\mbox{\tiny GR}}}
\newcommand{\B}{{\mbox{\tiny (B)}}}
\newcommand\dfz{f^{\prime}_0}
\newcommand\dfzs{f_0^{\prime 2}}
\newcommand\ddfz{f^{\prime \prime}_0}

\newcommand{\UVA}{Department of Physics, University of Virginia, Charlottesville, Virginia 22904-4714, USA}

\AtBeginDocument{%
    \newwrite\bibnotes
    \def\bibnotesext{Notes.bib}
    \immediate\openout\bibnotes=\jobname\bibnotesext
    \immediate\write\bibnotes{@CONTROL{REVTEX41Control}}
    \immediate\write\bibnotes{@CONTROL{%
    apsrev41Control,author="08",editor="1",pages="1",title="0",year="1"}}
    \if@filesw
    \immediate\write\@auxout{\string\citation{apsrev41Control}}%
    \fi
}

\begin{document}

\title{Analytic estimates of quasi-normal mode frequencies for \\ black holes in General Relativity and beyond}
\author{Kent Yagi}
\affiliation{\UVA}
\date{\today}

\begin{abstract}
In this chapter, we review the eikonal technique to analytically derive approximate quasi-normal mode frequencies of black holes. We first review the procedure in General Relativity and extend it to theories beyond General Relativity. As an example of the latter, we focus on scalar Gauss-Bonnet gravity in which a scalar field can couple to the Gauss-Bonnet invariant in an arbitrary way in the action. In this theory, metric and scalar perturbations are coupled in the polar sector, forcing the isospectrality to break and making the eikonal calculation more complex. We show that the analytic estimates can accurately capture the numerical behavior of the quasi-normal mode frequencies, especially when the coupling constant of the theory is small.

\end{abstract}

\maketitle

\tableofcontents

\newpage

\section{Introduction}
\label{sec:intro}

\begin{figure}
\includegraphics[width=8.cm]{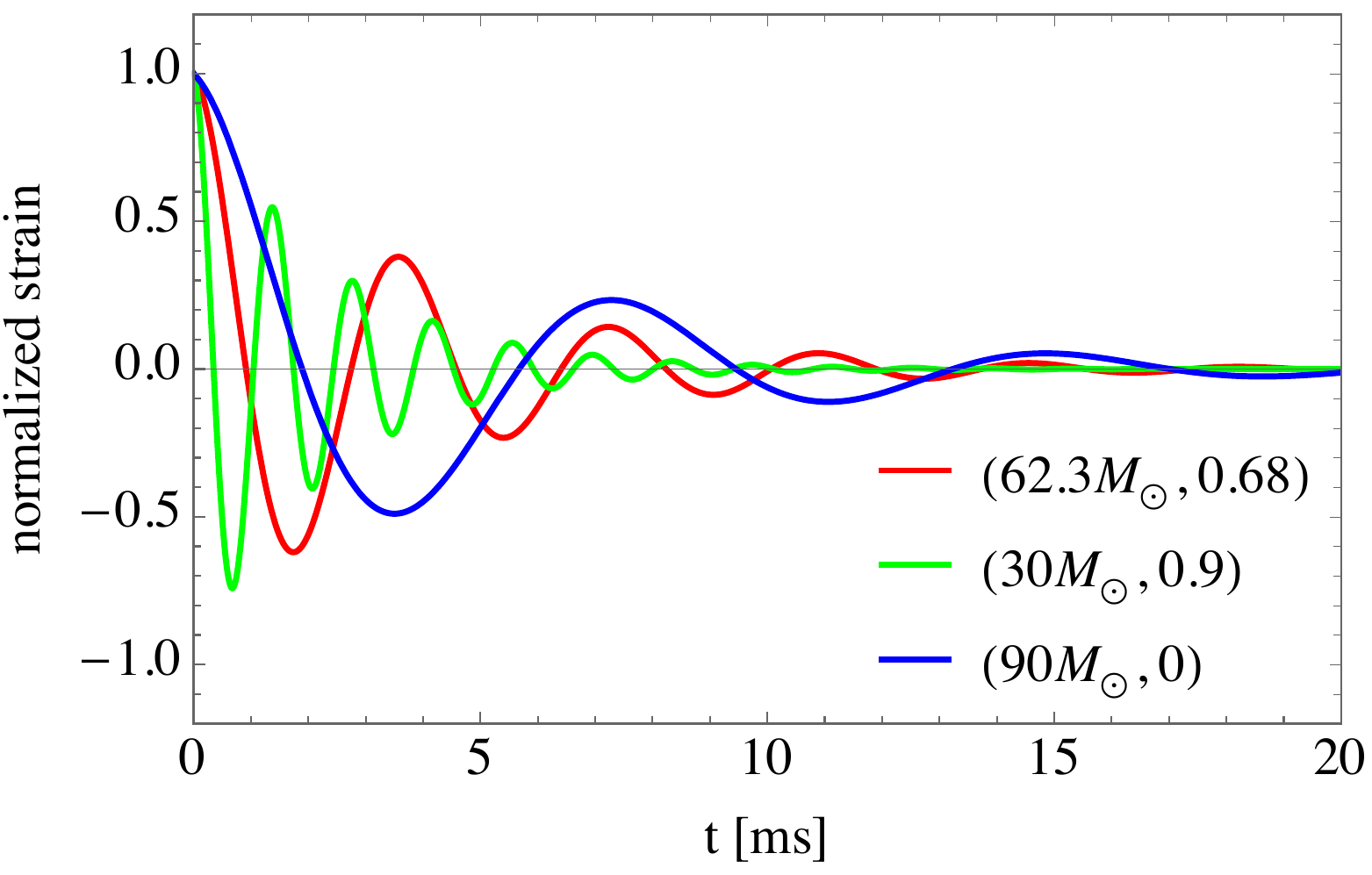}
\includegraphics[width=7.cm]{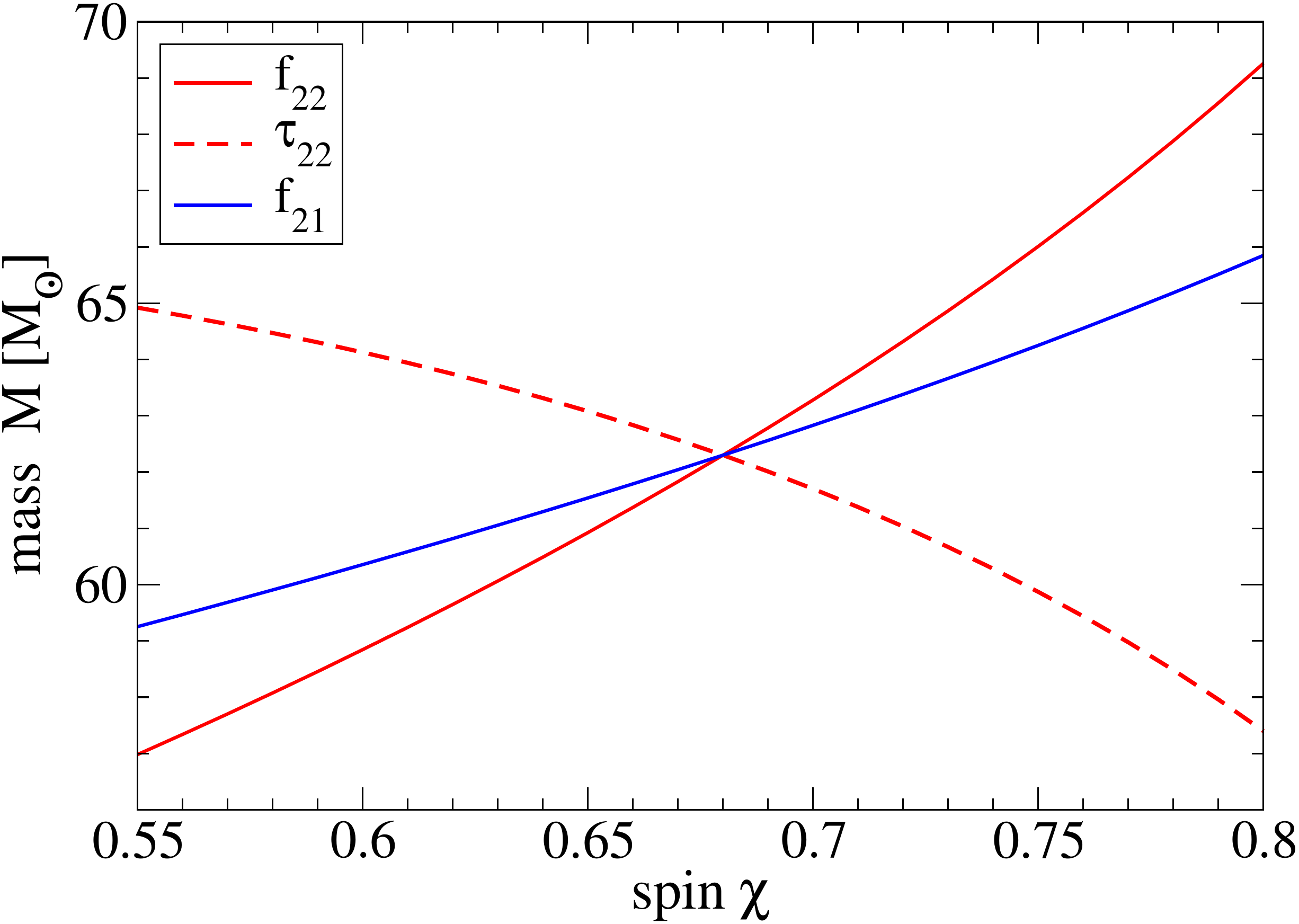}
\caption{\label{fig:ringdown-consistency}
(Left) Normalized GW strain of the BH ringdown in GR with $(\ell,m)=(2,2)$ of the form $h_{22}(t) = e^{-t/\tau_{22}} \cos(2\pi f_{22}t)$ for various BH mass and dimensionless spin combinations. The ringdown frequency $f_{22}$ and damping time $\tau_{22}$ are taken from the fit in~\cite{Berti:2005ys}.
(Right) Determination of the BH mass and spin from the ringdown frequency and damping time. We used the values of $f_{\ell m}$ and $\tau_{\ell m}$ to be the GR ones with $(M,\chi) = (62.3M_\odot,0.68)$ corresponding to GW150914~\cite{Abbott:2016blz}. GR is consistent if there is an overlap between different curves.
}
\end{figure}

A post-merger signal of gravitational waves (GWs) from a black hole (BH) coalescence is characterized by a quasi-normal mode (QNM) ringdown~\cite{Kokkotas:1999bd,Berti:2009kk} (see the left panel of Fig.~\ref{fig:ringdown-consistency}). Normal modes are a pattern of motion in which all parts of the system oscillate with the same frequency. For example, masses on springs will undergo normal mode oscillations if there is no friction. On the other hand, QNMs are normal modes with damping in a dissipative system. Namely, the oscillation is a sinusoid with its amplitude damping exponentially. If there is friction in the masses on springs mentioned above, they will undergo QNM oscillations. If one perturbs a BH spacetime, GWs would propagate to infinity or get absorbed to the BH. The system is dissipative and hence GWs follow QNM oscillations.

In General Relativity (GR), BHs possess a no-hair property~\cite{israel,hawking-uniqueness} and an astrophysical BH is characterized only by its mass and spin. Thus, complex QNM frequencies (or the frequency and damping time) are also characterized by the BH's mass and spin. The dominant GW mode has the harmonic of $(\ell,m)=(2,2)$. If one can measure the frequency and damping time of this dominant GW mode, one can determine the mass and spin of the remnant BH assuming GR is correct. If one can further measure the frequency or the damping time of a sub-leading mode (higher angular modes or overtones~\cite{JimenezForteza:2020cve}), one can check the consistency of the mass-spin measurement (see the right panel of Fig.~\ref{fig:ringdown-consistency}). If all curves in the mass-spin plane cross at a single point (or if there is a region in the parameter space where all curves with measurement errors overlap), the GR assumption is consistent with the observation. This way, one can probe gravity, known as the \emph{BH spectroscopy}~\cite{Dreyer:2003bv}. Such kinds of tests have already been performed with the existing GW events~\cite{Isi:2019aib,LIGOScientific:2020tif,Capano:2021etf} (see~\cite{Cotesta:2022pci} for some caution on distiunguishing overtone signals from noise). In principle, one could coherently stack signals from multiple events to enhance the detectability of the sub-leading modes~\cite{Yang:2017zxs}.

Although accurate estimates of the QNM frequencies require numerical calculations, there are a few analytic techniques available to derive approximate QNM frequencies. In this chapter, we focus on the \emph{eikonal} (or geometric optics) approximation~\cite{Press:1971wr,1972ApJ...172L..95G} where we assume the harmonic $\ell$ to be large ($\ell \gg 1$). In this eikonal picture, one can view the fundamental QNM as a wavepacket localized at the peak of the potential of the radial metric perturbation equation. Since the peak location of the potential coincides with the location of the photon ring within this approximation, complex QNM frequencies are associated with certain properties of null geodesics at the photon ring (orbital frequency and Lyapunov exponent)~\cite{Ferrari:1984zz,Mashhoon:1985cya,Cardoso:2008bp,Dolan:2010wr,Yang:2012he}.
We review how the eikonal calculation can be used to find BH QNM frequencies in GR and beyond~\cite{Glampedakis:2019dqh,Silva:2019scu,Bryant:2021xdh}. As an example of the latter, we consider scalar Gauss-Bonnet (sGB) gravity~\cite{Nojiri:2005vv,Yagi:2012gp,Antoniou:2017hxj,Antoniou_2018}, which is a generalization of Einstein-dilaton Gauss-Bonnet (EdGB) gravity motivated by string theory. We use the geometric unit of $c=G=1$ throughout. A prime of a function means taking the derivative with respect to its argument.

\section{General Relativity}

We start by reviewing the black hole perturbation equations and eikonal QNM calculations in GR. 

\subsection{Perturbation Equations}

We begin by considering a scalar perturbation under the Schwarzschild background following Chapter 12 of~\cite{Maggiore:2018sht}. We will then promote the perturbation equation to the tensor perturbation case. 

\begin{figure}
\includegraphics[width=11.5cm]{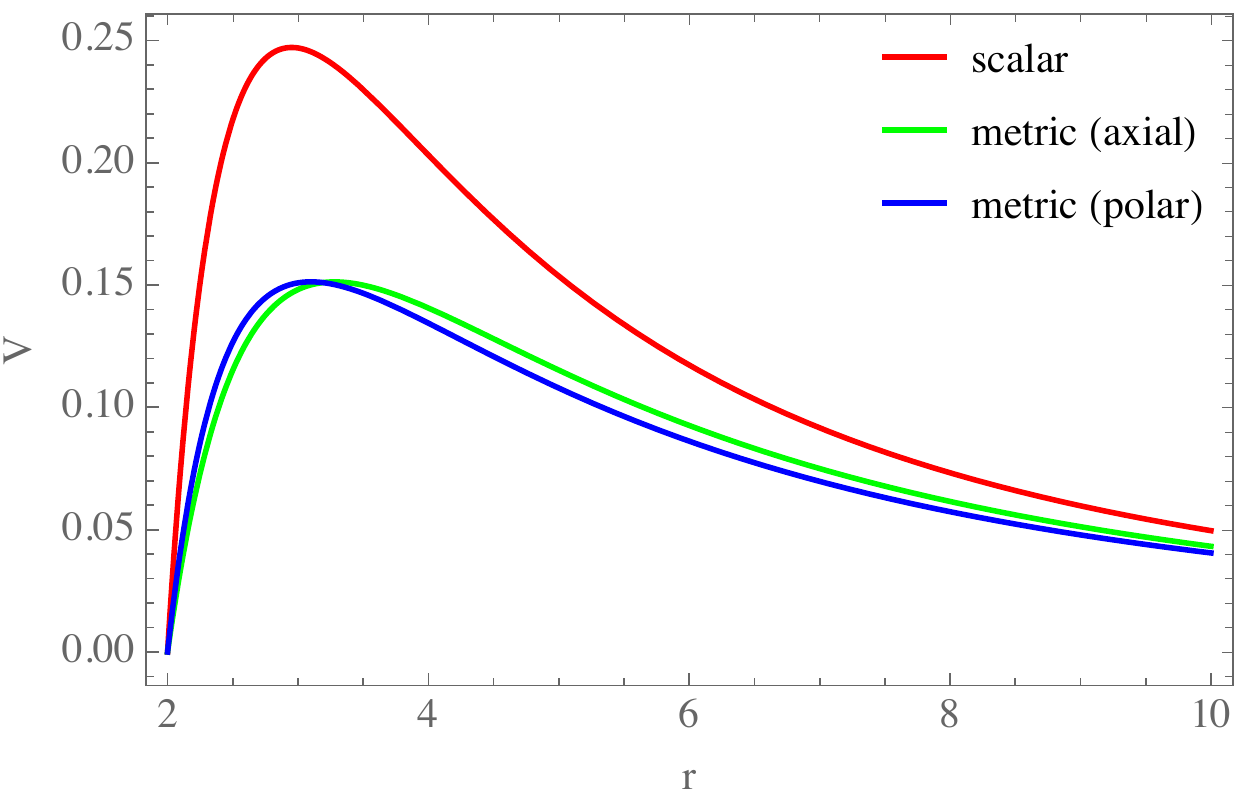}
\caption{\label{fig:pot_GR} Various potentials in GR with $\ell=2$ as a function of $r$ in the unit $M=1$. We show $V_2^\phi$ (red), $V_2^-$ (green) and $V_2^+$ (blue).
}
\end{figure}

Let us first derive the perturbation equation for the scalar field in real space.
The background metric $g^{\B}_{\mu\nu}$ is given by 
\begin{equation}
    ds^2 =  g^{\B}_{\mu\nu} dx^\mu dx^\nu =  - A_{\GR}(r) dt^2 + \frac{1}{A_{\GR}(r)}dr^2 + r^2 (d\theta^2 + r^2 \sin^2\theta d\phi^2)\,,
\end{equation}
where $A_{\GR} = 1-R_s/r$ with $R_s = 2M$ being the Schwarzschild radius for a BH with mass $M$. A test scalar field under this background metric follows the Klein-Gordon equation given by
\begin{equation}
\label{eq:KG}
\Box \phi = \frac{1}{\sqrt{-g^{\B}}} \partial_\mu \left( \sqrt{- g^{\B}} g_{\B}^{\mu\nu} \partial_\nu \right) \phi = 0\,,
\end{equation}
with $g^\B$ representing the metric determinant.
Since the background metric is spherically symmetric, one can expand the scalar field $\phi$ in terms of the spherical harmonics $Y_{\ell m}$ as
\begin{equation}
\label{eq:scalar_expand}
\phi(t,r,\theta,\phi) = \frac{1}{r} \sum_{\ell=0}^\infty \sum_{m=-\ell}^\ell \phi_{\ell m}(t,r) Y_{\ell m}(\theta,\phi)\,.
\end{equation}
Substituting Eq.~\eqref{eq:scalar_expand} into Eq.~\eqref{eq:KG}, one finds
\begin{equation}
\label{eq:scalar_pert_1}
A_{\GR} \partial_r (A_{\GR}\partial_r \phi_{\ell m}) - \partial_t^2 \phi_{\ell m} - V_\ell^\phi (r) \phi_{\ell m} = 0\,,
\end{equation}
with
\begin{equation}
V_\ell^\phi (r) = A_{\GR}(r) \left[\frac{\ell (\ell +1)}{r^2} + \frac{R_2}{r^3} \right]\,.
\end{equation}
We present $V_2^\phi(r)$ in Fig.~\ref{fig:pot_GR}.
Let us next move to a tortoise coordinate given by
\begin{equation}
x \equiv r + R_s \ln \frac{r-R_s}{R_s}\,.
\end{equation}
$x \to \infty$ corresponds to infinity while $x \to - \infty$ corresponds to the horizon.
Using $\partial_x r = A_{\GR}$ and $A_{\GR} \partial_r = \partial_x$, Eq.~\eqref{eq:scalar_pert_1} becomes
\begin{equation}
\label{eq:u}
\left[ \partial_x^2 - \partial_t^2 - V_\ell^\phi (r) \right] \phi_{\ell m}(t,r) = 0\,.
\end{equation}
Let us next look at Fourier modes. Substituting
\begin{equation}
\phi_{\ell m}(t,r) = \int^{\infty}_{-\infty} \frac{d\omega}{\sqrt{2\pi}} \tilde{\phi}_{\ell m}(\omega,r) e^{-i \omega t}\,,
\end{equation}
to Eq.~\eqref{eq:u}, we find
\begin{equation}
\label{eq:scalar_pert}
\left[ - \partial_x^2 + V_\ell^\phi(r) \right] \tilde \phi_{\ell m} = \omega^2 \tilde \phi_{\ell m}\,.
\end{equation}

So far, we have focused on a scalar perturbation (with spin $s=0$) under a Schwarzschild background, but the structure of the tensor perturbation equation is the same as in Eq.~\eqref{eq:scalar_pert}:
\begin{equation}
\label{eq:master_eq_GR}
\partial_x^2 \psi_{\pm} + [\omega^2 - V_\ell^{\pm}]\psi_\pm = 0\,,
\end{equation}
where the indices $+$ and $-$ refer to the polar (or even-parity) and axial (odd-parity) modes respectively and we drop the indices $\ell m$ for simplicity. Notice that these two perturbation modes decouple. For axial modes, $\psi_-$ is the Regge-Wheeler function (a linear combination of axial metric perturbation components) and the Regge-Wheeler potential $V_\ell^-$ is given by
\begin{equation}
V_\ell^- (r) \equiv A_{\GR}(r) \left[\frac{\ell (\ell +1)}{r^2} -3 \frac{R_s}{r^3} \right]\,.
\end{equation}  
On the other hand, for polar modes, $\psi_+$ is the Zerilli function (a linear combination of polar metric perturbation components) and the Zerilli potential $V_\ell^+$ is given by
\begin{equation}
V_\ell^+ (r) \equiv A_{\GR}(r) \frac{2 \left[\Lambda ^2 (\Lambda +1) r^3+3 \Lambda ^2 r^2+9 \Lambda  r+9\right]}{r^3
   (\Lambda  r+3)^2}\,,
\end{equation}
with $\Lambda = (\ell-1)(\ell+2)/2$. $V_2^\pm(r)$ are shown in Fig.~\ref{fig:pot_GR}. Observe that these potentials for metric perturbations are smaller than that for the scalar perturbation. Notice also that the peak locations are all at around $r=3M$ which is the location of the photon ring. 

We can understand the condition for deriving QNM frequencies by considering a scattering problem of waves under the above potential. The asymptotic behaviors of $\psi_\pm$ is given by
\begin{equation}
\psi_\pm\sim 
\begin{cases}
\mathcal A_i(\omega) e^{+i\omega x} + \mathcal A_r(\omega) e^{-i\omega x} & (x \to -\infty)\,, \\
\mathcal A_t(\omega) e^{+i\omega x}  &  (x \to +\infty)\,,
\end{cases}
\end{equation}
where $\mathcal A_i$, $\mathcal A_r$ and $\mathcal A_t$ are the amplitude for the initial, reflected and transmitted waves respectively. The condition to compute QNM is 
\begin{equation}
\mathcal A_i(\omega) = 0\,.
\end{equation}
Namely, we impose the wave to be purely ingoing at the horizon while purely outgoing at infinity.
Although the potentials for the axial and polar modes are different, QNM frequencies for these modes are identical. This intriguing property is called \emph{isospectrality}.

\subsection{Eikonal QNM Calculations}
\label{sec:eikonal_GR}

We now review analytic estimates for QNM frequencies in GR within the eikonal (high frequency) approximation. We refer the readers to e.g.~\cite{Glampedakis:2019dqh} for more details. Given the isospectrality, we focus on the axial modes. 

Eikonal approximation begins by making the following ansatz
\begin{equation}
\label{eq:eikonal_anz}
\psi_\pm = \mathcal A_\pm(x) e^{i S_\pm(x)/\epsilon}\,,
\end{equation}
for amplitude function $\mathcal A_\pm$ and phase function $S_\pm$ with $\epsilon \sim \ell^{-1} \ll 1$. Substituting Eq.~\eqref{eq:eikonal_anz} to Eq.~\eqref{eq:master_eq_GR}, we find
\begin{equation}
\label{eq:eikona_GR_2}
\partial_x^2 \mathcal A_\pm + \frac{i}{\epsilon}[2 (\partial_x S_\pm) (\partial_x \mathcal A_\pm) + \mathcal A_\pm \partial_x^2 S_\pm] + \left\{ \omega^2 - \frac{1}{\epsilon^2} (\partial_x^2 S_\pm)^2 - A_\GR \left[ \frac{\ell (\ell +1)}{r^2} -3 \frac{R_s}{r^3} \right]\right\}\mathcal A_\pm = 0\,.
\end{equation}
In the double expansion of $\epsilon \ll 1$ and $\ell \gg 1$, the above equation reduces to
\begin{equation}
\label{eq:eikonal_GR}
- \frac{1}{\epsilon^2} (\partial_x S_\pm)^2 + \omega^2 - \ell^2 U = 0\,,
 \end{equation}
to leading order with $U(r) \equiv A_\GR(r)/r^2$. Notice that $\omega = \mathcal{O}(\ell)$ since $\ell \sim 1/\epsilon$. 

For QNMs, we want $S_\pm(x) \to + \omega x$ for $x \to + \infty$ and $S_\pm(x) \to - \omega x$ for $x \to - \infty$, and $S_\pm$ to take its minimum value at the peak location $r=r_m = 3M$ at the leading eikonal order for the potential $U$. To check this, we can take the derivative of Eq.~\eqref{eq:eikonal_GR} with respect to $x$ to yield
\begin{equation}
\frac{2}{\epsilon^2}(\partial_x S_\pm) (\partial_x^2 S_\pm) = - \ell^2 \frac{dr}{dx} \frac{dU}{dr}\,.
\end{equation}
Thus, at the potential peak, $(dU/dr)_{m} = 0$ and $(\partial_x S_\pm)_{m} = 0$ where the subscript $m$ indicates that the quantity is evaluated at $r=r_m$. 

Let us now derive the real QNM frequency to leading eikonal order. We evaluate Eq.~\eqref{eq:eikonal_GR} at $r=r_m$ and impose $(\partial_x S_\pm)_{m} = 0$ to yield
\begin{equation}
\label{eq:omegaR0-GR}
\omega_R^{(0)} = \ell \sqrt{U_m} = \frac{\ell}{3\sqrt{3} M}\,,
\end{equation}
where the superscript (0) indicates that this frequency is the leading eikonal contribution.

As mentioned in Sec.~\ref{sec:intro}, there is a correspondence between QNM frequencies and null geodesic properties in GR. For example, the real part of the QNM frequency is approximately related to the orbital angular frequency $\Omega_\mathrm{ph}$ at the photon ring $r_\mathrm{ph}$ as 
\begin{equation}
\label{eq:geod_corr}
\omega_R = \ell \, \Omega_\mathrm{ph}\,,
\end{equation}
where
\begin{equation}
\label{eq:Omega_ph}
\Omega_\mathrm{ph} = \frac{\sqrt{A_\GR(r_\mathrm{ph})}}{r_\mathrm{ph}}\,.
\end{equation}
The photon ring location is obtained by solving
\begin{equation}
2 A_\GR(r_\mathrm{ph}) = r_\mathrm{ph}A_\GR'(r_\mathrm{ph})\,,
\end{equation}
to yield $r_\mathrm{ph}=3M$ (which coincides with the potential peak location $r_m$ in the eikonal limit). Plugging this into Eq.~\eqref{eq:Omega_ph}, we find $\Omega_\mathrm{ph} = 1/(3\sqrt{3}M)$ and Eq.~\eqref{eq:geod_corr} agrees with Eq.~\eqref{eq:omegaR0-GR}.

To find the imaginary QNM frequency, we need to go to the next-to-leading eikonal order. For $\omega = \omega_R + i \omega_I$ with $\omega_R, \omega_I \in \mathbb{R}$, we make the following ansatz:
\begin{equation}
\label{eq:ansatz_omega}
\omega_R = \omega_R^{(0)} + \omega_R^{(1)} + \mathcal{O}(\ell^{-1})\,, \quad \omega_I = \omega_I^{(1)} + \mathcal{O}(\ell^{-1})\,,
\end{equation}
where the superscript $(N)$ indicates the quantity to be at $N$th order in the eikonal expansion or $\mathcal{O}(\ell^{1-N})$. Substituting this to Eq.~\eqref{eq:eikona_GR_2}, expand in $\epsilon \ll 1$ and $\ell \gg 1$ and keeping only the contribution at $\mathcal{O}(\ell, \epsilon^{-1})$, we find
\begin{equation}
\label{eq:eikonal_GR_3}
\frac{2i}{\epsilon}(\partial_x S_\pm) (\partial_x \mathcal A_\pm) + \left( \frac{i}{\epsilon}\partial_x^2 S_\pm + 2 i \omega_R^{(0)} \omega_I^{(1)} + 2 \omega_R^{(0)} \omega_R^{(1)} - \ell U\right) \mathcal A_\pm = 0\,.
\end{equation} 

To proceed further, we need $(\partial_x^2 S_\pm)_{m}$. To find this, we Taylor expand Eq.~\eqref{eq:eikonal_GR} about $r=r_m$ to yield
\begin{equation}
\frac{1}{\epsilon^2}(\partial_x S_\pm)^2
 = - \frac{\ell^2}{2} U_m''(r-r_m)^2\,.
 \end{equation}
Taking the positive root for $\partial_x S_\pm$ and differentiate with respect to $x$, we find
\begin{equation}
\label{eq:ddS}
\frac{1}{\epsilon}(\partial_x^2 S_\pm) = \frac{\ell}{\sqrt{2}} \frac{dr}{dx} \sqrt{|U_m''|}\,.
\end{equation}

We are now ready to evaluate the imaginary QNM frequency and the sub-leading QNM real frequency. Regarding the former, we take the imaginary part of Eq.~\eqref{eq:eikonal_GR_3}, together with Eqs.~\eqref{eq:omegaR0-GR} and~\eqref{eq:ddS}, and evaluate it at $r_0$ to find
\begin{equation}
\omega_I^{(1)} = - \frac{1}{2\epsilon} \frac{(\partial_x^2 S_\pm)_{m}}{\omega_R^{(0)}} = - \frac{1}{2} \left( \frac{dr}{dx} \right)_m \sqrt{\frac{|U_m''|}{2 U_m}} = - \frac{1}{6\sqrt{3}M}\,.
\end{equation}

Regarding the latter, we take the real part of Eq.~\eqref{eq:eikonal_GR_3} to find 
\begin{equation}
\omega_R^{(1)} =\frac{\ell U_m}{2 \omega_R^{(0)}} = \frac{1}{2} \sqrt{U_m} = \frac{1}{6\sqrt{3}M}\,.
\end{equation}
Together with $\omega_R^{(0)}$, we find
\begin{equation}
\label{eq:eikonal_real_f}
\omega_R = \left( \ell + \frac{1}{2} \right) \sqrt{U_m} = \frac{1}{3\sqrt{3}M}\left( \ell + \frac{1}{2} \right)\,. 
\end{equation}

\begin{figure}
\includegraphics[width=11.5cm]{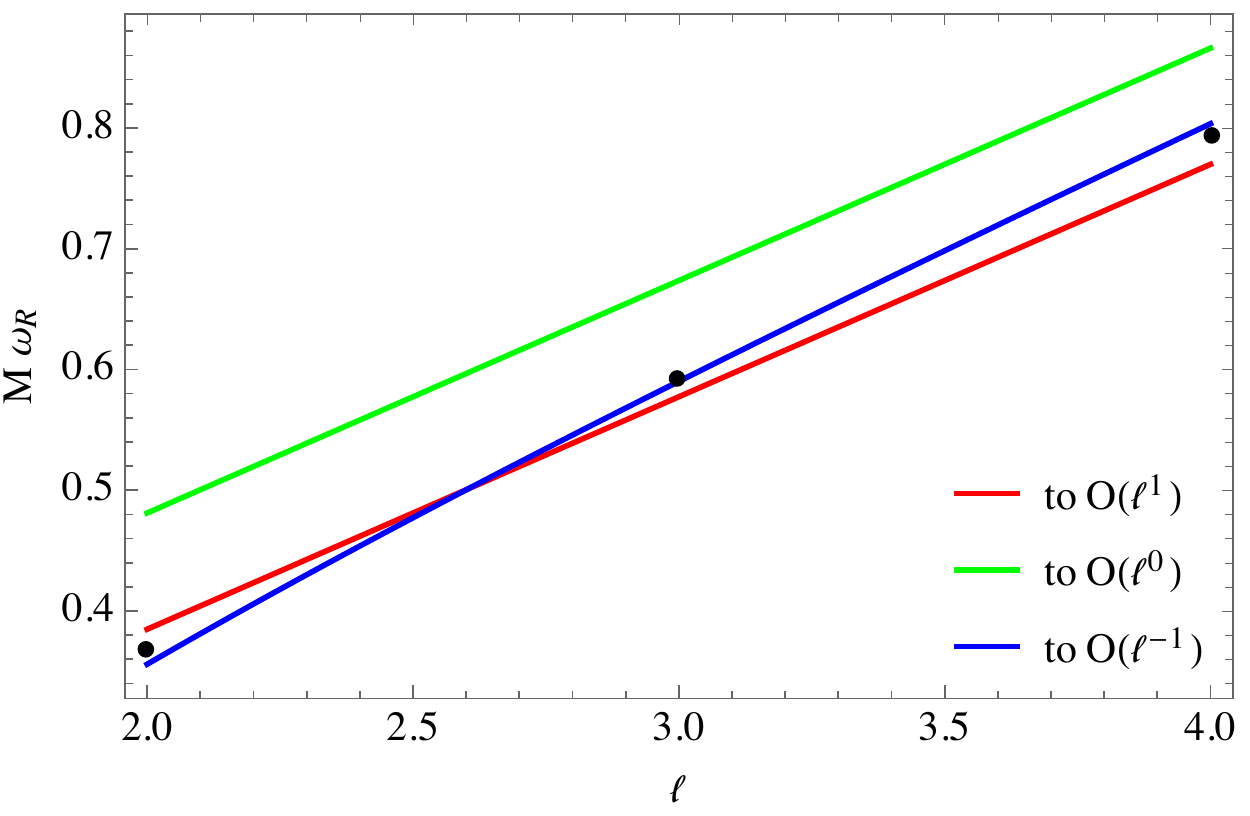}
\caption{\label{fig:comparison} Comparison of the analytic estimate for the real QNM frequency at various $\ell$ in GR with the numerical values~\cite{Berti:2005ys} (black dots). For the former, we present the result with the leading eikonal contribution (red), and including the next-to-leading contribution (green) in Eq.~\eqref{eq:eikonal_real_f}. We also present the result including the next-to-next-to-leading contribution taken from~\cite{Iyer:1986nq} through the WKB approximation.
}
\end{figure}

Figure~\ref{fig:comparison} compares the above eikonal results with the numerical values at various $\ell$. The leading eikonal results at $\mathcal{O}(\ell)$ agrees with the numerical ones within an error of $4.5 \%$. Interestingly, the agreement becomes worse if we include the next-to-leading contribution at $\mathcal{O}(\ell^0)$. This problem is cured by including the next-to-next-to-leading contribution at $\mathcal{O}(\ell^{-1})$ found in~\cite{Iyer:1986nq} via a WKB approximation. Now the two results agree within an error of $3.4\%$.

\section{Scalar Gauss-Bonnet Gravity}

We now review the application of the eikonal technique to theories beyond GR. 
We mainly review the work in~\cite{Bryant:2021xdh} which applies the eikonal analysis to another higher curvature theory called sGB gravity. 

\subsection{Theory and Black Hole Solution}

The action for this theory in vacuum is given by~\cite{Nojiri:2005vv,Yagi:2012gp,Antoniou:2017hxj,Antoniou_2018}
\begin{equation}
S = \frac{1}{16\pi} \int d^4 x \sqrt{-g} \left[R - \frac{1}{2}\partial_\mu \phi \partial^\mu \phi + \alpha f(\phi) \mathcal{G} \right]\,,
\end{equation}
where $\phi$ is the scalar (dilaton) field, $R$ is the Ricci scalar, $\alpha$ is the coupling constant, $f$ is an arbitrary function of $\phi$ and $\mathcal{G}$ is given by
\begin{equation}
\mathcal{G} = R_{\alpha\beta\mu\nu}R^{\alpha\beta\mu\nu} - 4R_{\alpha\beta}R^{\alpha\beta} + R^2\,,
\end{equation}
with $R_{\alpha\beta\mu\nu}$ and $R_{\alpha\beta}$ being the Riemann and Ricci tensors respectively. $f(\phi) \propto \exp(\gamma \phi)$ for a constant $\gamma$ corresponds to EdGB gravity while $f(\phi) \propto \phi$ corresponds to shift-symmetric sGB gravity. SGB gravity and its extension is motivated also by cosmology, such as inflation~\cite{Oikonomou:2020sij,Oikonomou:2021kql}. The coupling constant $\alpha$ has been constrained by various observations, including low-mass x-ray binaries~\cite{Yagi:2012gp}, gravitational waves from binary black holes~\cite{Nair:2019iur,Yamada:2019zrb,Tahura:2019dgr,Perkins:2021mhb,EdGB_Wang,Lyu:2022gdr,Perkins:2022fhr} and neutron star observations~\cite{Pani:2011xm,Saffer:2021gak} (see also a recent work in~\cite{Herrero-Valea:2021dry}). 

The field equations are given by
\begin{align}
\Box\phi &= \alpha f'(\phi) \mathcal G\,,  \\
%
G_{\alpha\beta} &= \frac{1}{2}\partial_\alpha\phi\,\partial_\beta\phi-\frac{1}{4}g_{\alpha\beta}\partial_\mu\phi\, \partial^\mu\phi-\alpha\mathcal{K}_{\alpha\beta}\,,
\end{align}
where $G_{\alpha\beta}$ is the Einstein tensor while $\mathcal{K}_{\alpha\beta}$ is given by
\begin{equation}
\mathcal{K}_{\alpha\beta}=(g_{\alpha\mu}g_{\beta\nu}+g_{\alpha\nu}g_{\beta\mu}) \, 
\epsilon^{\rho\nu \sigma \kappa}\nabla_\lambda \,
[{}^{\ast}{R}^{\mu \lambda}{}_{\sigma \kappa}\partial_\rho f(\phi)]\,,
\end{equation}
with the dual Riemann tensor ${}^*R^{\alpha\beta}{}_{\gamma\delta} = \epsilon^{\alpha\beta\mu\nu} R_{\mu\nu\gamma\delta}$ and $\epsilon^{\alpha\beta\mu\nu}$ representing the Levi-Civita tensor.

In this theory, a non-rotating black hole solution is known analytically within the small coupling approximation ($\alpha \ll M^2$). To $\mathcal{O}(\alpha^2)$, the metric is given by~\cite{Kanti:1995vq,Yunes:2011we,Pani:2011gy,Ayzenberg:2014aka,Maselli:2015tta}
\begin{equation}
ds^2 = \bar g^\B_{\mu\nu} dx^\mu dx^\nu =  - A(r) dt^2 + \frac{1}{B(r)}dr^2 + r^2 (d\theta^2 + r^2 \sin^2\theta d\phi^2)\,,
\end{equation}
with
\begin{align}
A &= 1-\frac{2M}{r} - \frac{\alpha^2 \dfzs}{r M^3} 
\left(
\frac{49}{40} 
- \frac{M^2}{3 r^2}
- \frac{26M^3}{3 r^3}
- \frac{22M^4}{5 r^4}
- \frac{32M^5}{5 r^5}
+ \frac{80M^6}{3 r^6}
\right)
\,, 
\\
B &= 1-\frac{2M}{r} - \frac{\alpha^2 \dfzs}{rM^3} 
\left(
\frac{49}{40}
- \frac{M}{r}
- \frac{M^2}{r^2}
- \frac{52M^3}{3 r^3}
- \frac{2M^4}{r^4}
- \frac{16M^5}{5 r^5}
+ \frac{368M^6}{3 r^6}
\right)\,,
\end{align}
and $f'_0 \equiv f'(0)$. The background scalar field is given by
\begin{equation}
\phi_0 = \frac{2 \alpha \dfz}{r M}
\left(
1 
+ \frac{M}{r} 
+ \frac{4M^2}{3 r^2}
\right)
+ \frac{\alpha^2 \dfz \ddfz}{rM^3}
\left(
\frac{73}{30}
+ \frac{73M}{30 r}
+ \frac{146M^2}{45 r^2}
+ \frac{73M^3}{15 r^3}
+ \frac{224M^4}{75 r^4}
+ \frac{16M^5}{9 r^5}
\right)\,,
\end{equation}
with $f''_0 \equiv f''(0)$.
The ADM mass $M_*$ is given in terms of the GR one $M$ as follows:
\begin{equation}
\label{eq:ADM}
M_* = M\left(1 + \frac{49}{80} \frac{\alpha^2 \dfzs}{M^4}\right)\,.
\end{equation}

\subsection{Perturbation Equations}

Next, we look at black hole perturbation equations in sGB gravity. We perturb both the metric and scalar field as
\begin{equation}
g_{\mu\nu}= \bar g^\B_{\mu\nu} + h_{\mu\nu}\,, \quad \phi = \phi_0 + \delta \phi\,,
\end{equation}
where $h_{\mu\nu}$ and $\delta \phi$ are the metric and scalar perturbations respectively. As we will see below, $\delta \phi$ is coupled only to the polar sector of the metric perturbation, while the axial part is decoupled.

Let us begin with the axial perturbations. The master perturbation equation is given by~\cite{Blazquez-Salcedo:2016enn,Bryant:2021xdh}
\begin{equation}
\label{eq:axial_sGB}
\partial_{\bar x}^2 \bar \psi_- +(C_- \omega^2 - \bar V_\ell^-)\bar \psi_- = 0\,.
\end{equation}
Here $\bar x$ is the tortoise coordinate in sGB gravity that satisfies $\partial_r \bar x = 1/\sqrt{A B}$. $\bar \psi_-$ is the master variable for axial perturbation in sGB gravity, while $\bar V_\ell^-$ is the corresponding potential whose lengthy expression can be found in~\cite{Bryant:2021xdh}. The coefficient $C_-$ is given by
\begin{equation}
C_{-} =\frac{A}{A - 2 \alpha  B A' \phi_0' f_0'} 
\{ 1-2 \alpha  B' \phi_0' f_0'
 + 4 \alpha  B [{\phi_0'}^2 f_0'' + \phi_0''
   f_0'] \}\,.
\end{equation}
In the limit $\alpha \to 0$, $C_- \to 1$ while $\bar V_\ell^- \to V_\ell^-$ and Eq.~\eqref{eq:axial_sGB} reduces to the GR one in Eq.~\eqref{eq:master_eq_GR}. We can further perform a coordinate transformation from $\bar x$ to $\tilde x$ to remove $C_-$:
\begin{equation}
\frac{d \tilde x}{d \bar x} = \sqrt{C_-}\,.
\end{equation} 
Equation~\eqref{eq:axial_sGB} then becomes
\begin{equation}
\label{eq:axial_eq_sGB}
\partial_{\tilde x}^2 \bar \psi_- + p_- \partial_{\tilde x} \bar \psi_- + (\omega^2 - \tilde V_\ell^-) \bar \psi_- = 0\,,
\end{equation}
where $p_- \equiv \partial_{\bar x} C_-/(2 C_-)^{3/2}$ and $\tilde V_\ell^- \equiv \bar V_\ell^-/C_-$.

\begin{figure}
\includegraphics[width=11.5cm]{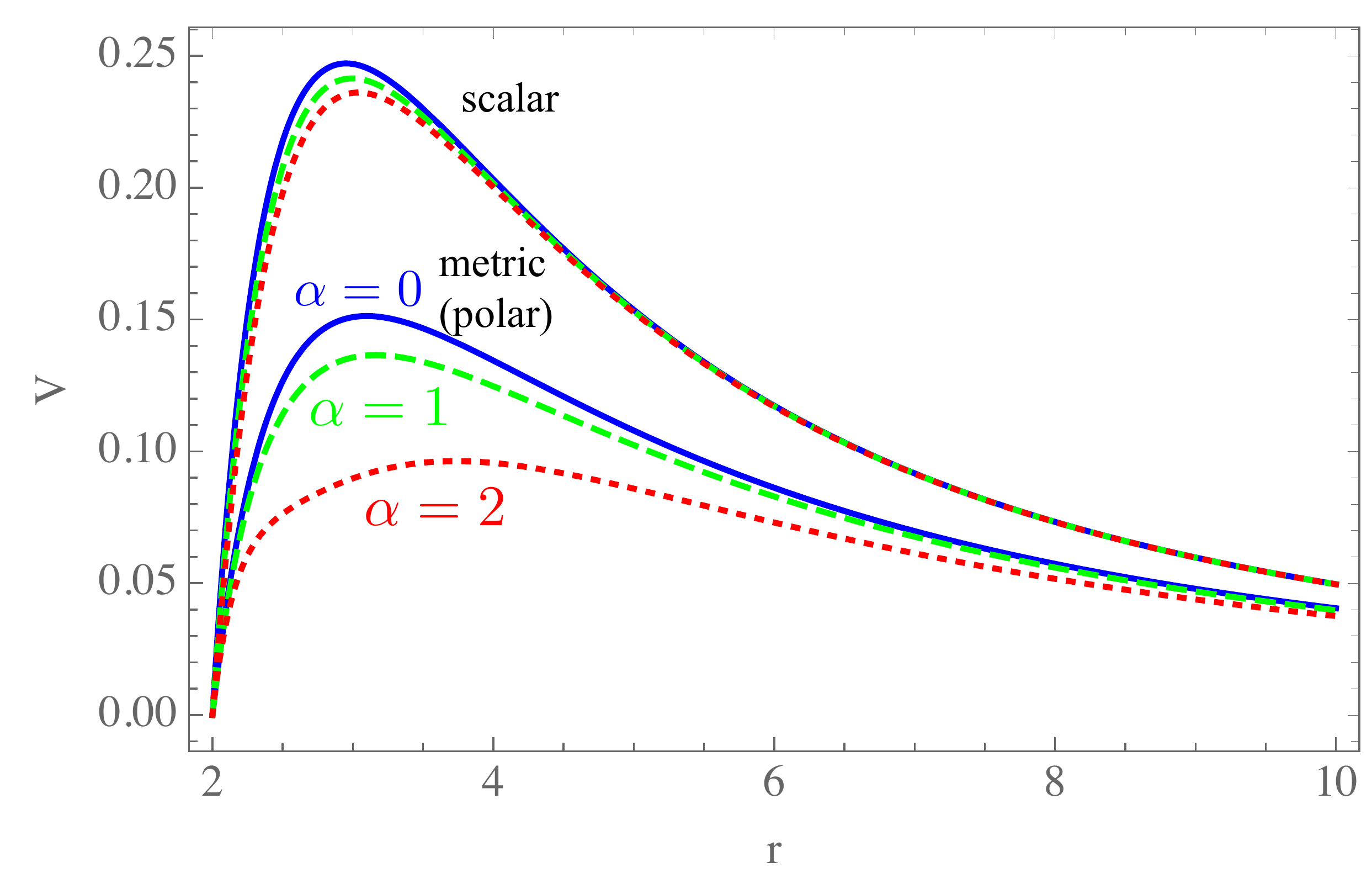}
\caption{\label{fig:pot_polar} Scalar potential $\bar V_2^\phi$ (top) and metric polar potential $\bar V_2^+$ (bottom) as a function of $r$ for three different choices of $\alpha$ (in the unit $M=1$).
}
\end{figure}

Next, we look at polar perturbations. The master equations for the metric and scalar perturbations under the small coupling approximation are given by~\cite{Bryant:2021xdh}
\begin{align}
\label{eq:polar_metric_eq_EdGB}
\partial_{\bar x}^2 \bar \psi_+ + p_+ \partial_{\bar x} \bar \psi_+ + (C_+ \omega^2 - \bar V_\ell^+) \bar \psi_+ = a_0 \bar \phi + a_1 \partial_{\bar x} \bar \phi\,, \\
\label{eq:polar_scalar_eq_EdGB}
\partial_{\bar x}^2 \bar \phi + (\omega^2 - \bar V_\ell^\phi) \bar \phi = b_0 \bar \psi_+ + b_1 \partial_{\bar x}\bar \psi_+\,.
\end{align}
Here $\bar \psi_+$ and $\bar \phi$ are the master metric and scalar perturbation variables respectively. 
For example,  $\bar \phi$ is given by
\begin{equation}
\delta \phi (t,r)  = \frac{1}{\sqrt{2\pi}} \int d t \, 
\frac{\bar{\phi}(\omega, r)}{r} Y_{\ell m} \, 
e^{-i \omega t}\,.
\end{equation}
The scalar potential is given by~\cite{Bryant:2021xdh}
\begin{equation}
\bar V_\ell^\phi = V_\ell^\phi -\alpha f_0'' \frac{48 M^2  (r-2M)}{r^7}\,,
\end{equation}
while the polar metric potential $\bar V_\ell^+$ and other coefficients in Eqs.~\eqref{eq:polar_metric_eq_EdGB} and~\eqref{eq:polar_scalar_eq_EdGB} are given in~\cite{Bryant:2021xdh}. Notice that the two perturbation equations are coupled. Figure~\ref{fig:pot_polar} presents $\bar V_2^\phi$ and $\bar V_2^+$ for various $\alpha$. Notice that the potential decreases while the peak location increases as one increases $\alpha$.

\subsection{Eikonal QNM Calculations}

Let us now apply the eikonal technique to find QNM frequencies in sGB gravity~
\cite{Bryant:2021xdh}. We will look at axial and polar sectors separately.

\subsubsection{Axial Perturbations}

We begin by studying the axial perturbations. The procedure is the same as that in GR described in Sec.~\ref{sec:eikonal_GR}. We start by making an ansatz
\begin{equation}
\bar \psi_- = \bar{\mathcal A}_-(\tilde x) e^{i \bar S_- (\tilde x)/\epsilon}\,,
\end{equation}
with the amplitude function $\bar{\mathcal A}_-$ and the phase function $\bar S_-$. Substituting this to Eq.~\eqref{eq:axial_eq_sGB}, expand in $\epsilon \ll 1$ and $\ell \gg 1$ and dropping $\partial_{\tilde x} \bar S_-$, we find
\begin{equation}
\label{eq:omega_R_axial_sGB_0}
\omega^{(0)}_R = \ell \sqrt{\bar U_m}=\ell \, \left[\frac{A-2 \alpha  B A' \phi_0' \dfz}{r \left(r - 4 \alpha 
   B \phi_0' \dfz\right)}\right]^{1/2}_{m} \,,
\end{equation}
where $\bar U$ is given by
\begin{equation}
\bar V_\ell^- = \ell^2 \bar U + \mathcal{O}(\ell)\,.
\end{equation}
We also need to account for the sGB correction to the peak location of the potential. Solving $\bar U'=0$, we find
\begin{equation}
r_m = 3M + \frac{6577}{19440}\frac{\alpha^2 \dfzs}{M^3}\,,
\end{equation}
to $\mathcal{O}(\alpha^2)$. Thus, under the small coupling approximation, we find
\begin{equation}
\label{eq:omega_R_sGB_small_coupling_0}
\omega_R^{(0)}=\frac{\ell}{3 \sqrt{3}M}\left(1- \frac{71987}{174960}\frac{\alpha ^2 \dfzs}{M^4} \right)\,.
\end{equation}

We comment on the correspondence between QNM and null geodesics. Notice that Eq.~\eqref{eq:omega_R_axial_sGB_0} is different from 
\begin{equation}
\label{eq:geod_corr_sGB}
\omega_R = \ell \frac{\sqrt{A(r_\mathrm{ph})}}{r_\mathrm{ph}}
\end{equation}
obtained by combining Eqs.~\eqref{eq:geod_corr} and~\eqref{eq:Omega_ph}. In particular, Eq.~\eqref{eq:omega_R_axial_sGB_0} depends on the background scalar field. Thus, the correspondence does not work in non-GR theories in general. However, for sGB gravity up to $\mathcal{O}(\alpha^2)$, the contribution from the background scalar field cancels with the correction to $r_\mathrm{ph}$ and one does recover Eq.~\eqref{eq:omega_R_sGB_small_coupling_0} by using Eq.~\eqref{eq:geod_corr_sGB} with $r_\mathrm{ph} = 3M$. Thus, the null geodesic correspondence still holds, at least up to $\mathcal{O}(\alpha^2)$ in the axial sector.

We now move to finding the sub-leading contribution to the axial QNM frequencies. We make the same ansatz to $\omega$ as in Eq.~\eqref{eq:ansatz_omega}. The only difference from the GR case is to replace $U$ by $\bar U$ and $x$ to $\tilde x$. For the real frequency, we find $\omega_R^{(1)} =\sqrt{\bar U_m}/2$ and
\begin{align}
\label{eq:omega_R_axial_sGB}
\omega_R = \left( \ell + \frac{1}{2} \right) \sqrt{\bar U_m} = &\left(  \ell+\frac{1}{2}\right) 
\left [\, \frac{A-2 \alpha  B A' \phi_0' \dfz}{r\left( r -4 \alpha 
 B \phi_0' \dfz \,\right)} \,\right]^{1/2}_m
 \nonumber \\
 =& \frac{1}{3 \sqrt{3}M}\left( \ell + \frac{1}{2} \right) \left(1- \frac{71987}{174960}\frac{\alpha ^2 \dfzs}{M^4} \right)\,.
\end{align}
For the imaginary frequency, we find
\begin{equation}
\label{eq:omega_I_axial_sGB}
\omega_I^{(1)}  = - \frac{1}{2} \left( \frac{dr}{d\tilde x} \right)_m \sqrt{\frac{|\bar U_m''|}{2 \bar U_m}} = -\frac{1}{6\sqrt{3}M}\left(1-\frac{121907}{174960}\frac{\alpha^2 \dfzs}{M^4} \right)\,,
\end{equation}
under the small coupling approximation.

\subsubsection{Polar Perturbations}

We next turn our attention to the polar sector. This time, the situation is quite different from the GR case as the perturbation equations for the metric and scalar fields are coupled. Our starting point is the same by making the following ansatz:
\begin{equation}
\bar \psi_+ = \bar{\mathcal A}_+(\tilde x) e^{i \bar S_+ (\tilde x)/\epsilon}\,, \quad \bar \phi = \bar{\mathcal A}_\phi(\tilde x) e^{i \bar S_+ (\tilde x)/\epsilon}\,.
\end{equation}
Notice that the phase functions are common in the two perturbations. Using this, we find that the equation for $\omega$ is biquadratic instead of quadratic:
\begin{equation}
\omega^4 + F(\epsilon,\alpha, \ell,r_m,\bar S_{+,m}, \bar{\mathcal{A}}_{+,m},\bar{\mathcal{A}}_{\phi,m}) \, \omega^2 
+ G(\epsilon,\alpha,\ell,r_m,\bar S_{+,m}, \bar{\mathcal{A}}_{+,m},\bar{\mathcal{A}}_{\phi,m}) = 0\,. 
\end{equation}
Working in the small coupling assumption of $\alpha \ll \epsilon$ (and setting the book-keeping parameter $\epsilon$ to 1), we find the solution to $\omega^2$ as
\begin{equation}
\omega_\pm^2 = \frac{\ell^2}{27M^2} \left \{\, 1 
+ \frac{1}{\ell} \left (1- 27 i\frac{S^{\prime\prime}_{+,m}}{\ell} \right )
\pm \frac{8\,\alpha^2  \dfzs{}}{27M^4} 
\left [\,\ell^2+ 2 \left (\ell -4 i S_{+,m}^{\prime\prime} \right )\,\right ]
\,\right\}\,,
\end{equation}
with
\begin{equation}
S_{+,m}^{\prime\prime}  = \frac{\ell}{27M^2} \left(
    1 - \frac{560}{2187} \frac{\alpha^2 \ell \dfzs}{M^4}\right)\,.
\end{equation}
Notice that there are two solutions for $\omega^2$. The mode with the $+(-)$ sign corresponds to the scalar-led (gravity-led) mode~\cite{Blazquez-Salcedo:2016enn}. From the above equations, the leading eikonal contribution to the real and imaginary frequencies are given by
\begin{align}
\omega_{R \pm} =& \frac{\ell}{3\sqrt{3}M} \left (\, 1  
\pm \frac{4}{27} \frac{\alpha^2 \ell^4 \dfzs{}}{M^4} \,\right )\,, \\
 \omega_{I\pm} =& - \frac{1}{6 \sqrt{3}M} \left( 1 \pm \frac{44}{729} \frac{\alpha^2 \ell^2 \dfzs}{M^4} \right)\,.
\end{align}
Comparing these with the axial results (Eqs.~\eqref{eq:omega_R_axial_sGB} and~\eqref{eq:omega_I_axial_sGB}), it is clear that the isospectrality is broken in sGB gravity. Moreover, there are three independent QNM frequencies: axial (gravity-led) mode, polar gravity-led mode, and polar scalar-led mode.

\subsection{Summary of Eikonal Expressions}

We end this section by summarizing the eikonal expressions and comparing them with the numerical results in~\cite{Blazquez-Salcedo:2016enn}. We convert the expressions in terms of the ADM mass $M_*$ using Eq.~\eqref{eq:ADM}.

\subsubsection{Axial Perturbations}

The axial QNM frequencies are given as follows:
\begin{align}
\omega_{R}&=\left(\ell+\frac{1 }{2}\right)\frac{1}{3\sqrt{3}M_* }\left(1+\frac{4397}{21870}
  \frac{\alpha^2 \dfzs}{M_*^4}\right), \\
\omega_{I}&=-\frac{1 }{6\sqrt{3}M_* }\left(1-\frac{1843}{21870}\frac{\alpha^2 \dfzs}{M_*^4}\right).
\end{align}

\subsubsection{Polar Perturbations}

For polar perturbations, the real frequency of the gravity-led mode is given by
\begin{equation}
\omega_{R -} = \frac{\ell}{3\sqrt{3}M_*} \left (\, 1  
- \frac{4}{27} \frac{\alpha^2 \ell^4 \dfzs{}}{M_*^4} \,\right )\,. 
\end{equation}
For the scalar-led mode, it turns out that the leading eikonal results are not sufficient to accurately describe the numerical results. For this reason, we include the higher order contributions:
\begin{equation}
\omega_{R+}
= \frac{\sqrt{\ell  (\ell +1 )}}{3 \sqrt{3}M_*} \left[ 1 - \frac{8\alpha}{27\ell  (\ell
   +1 )M_*^2}\ddfz  +\frac{4}{27}\frac{\alpha^2  \dfzs}{(\ell +1)M_*^4} \left(\ell^3+2 \ell^2+\frac{5249 \ell}{960}+\frac{1323}{320}\right) \right]\,.
\end{equation}

The imaginary frequencies for the gravity-led and scalar-led modes are given by
\begin{equation}
 \omega_{I\pm} = - \frac{1}{6 \sqrt{3}M_*} \left( 1 \pm \frac{44}{729} \frac{\alpha^2 \ell^2 \ddfz}{M_*^4} \right)\,.
\end{equation}

\subsubsection{Comparison with Numerical Results}

\begin{figure}
\includegraphics[width=11.5cm]{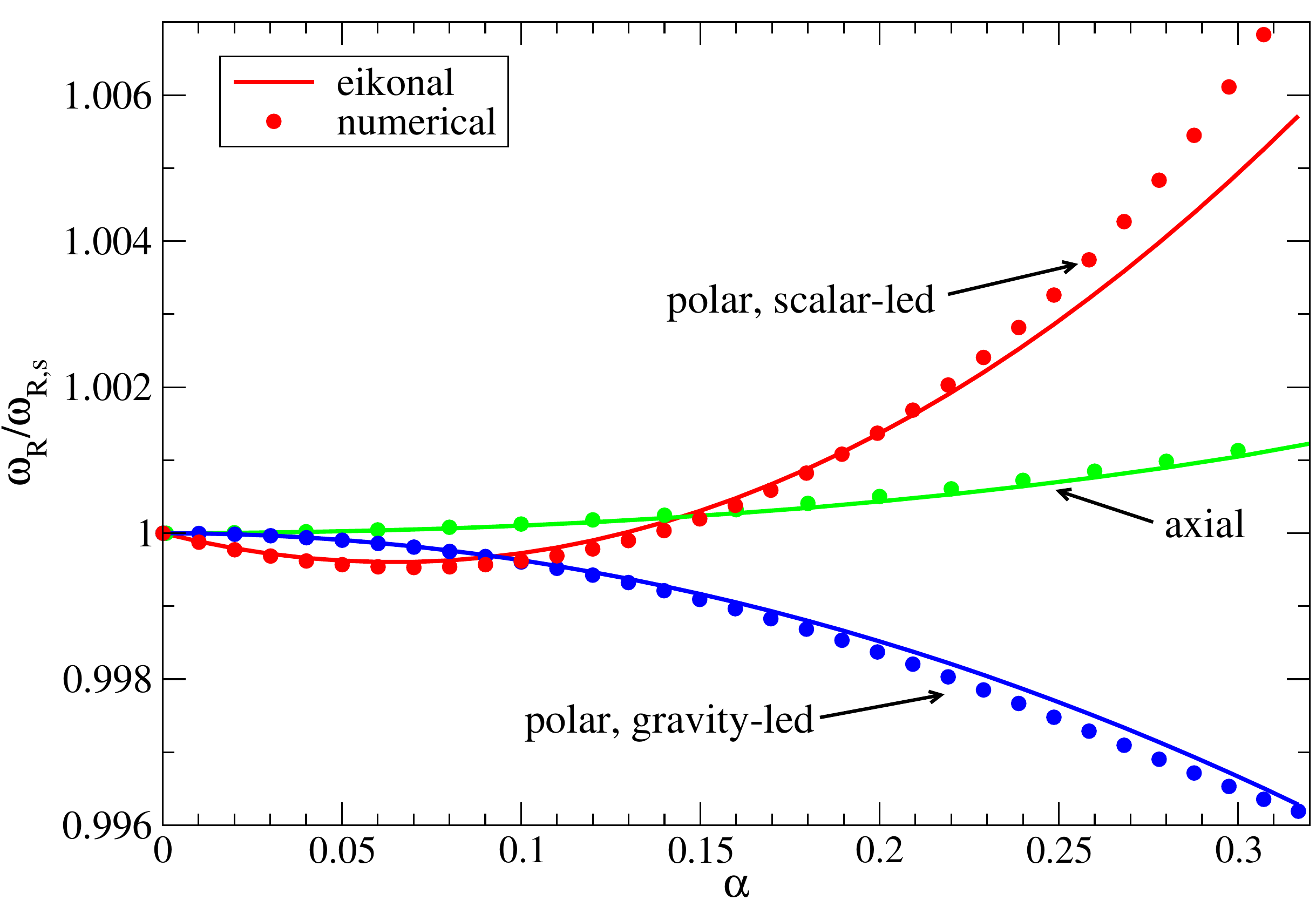}
\caption{\label{fig:summary}
Various $\ell=2$ real QNM frequencies in EdGB gravity with $f_0'=1/4$~\cite{Bryant:2021xdh}. We present the frequency normalized by the GR Schwarzschild value $(\omega_{R,S})$ as a function of the coupling constant $\alpha$ (in the unit $M=1$) for the analytic eikonal (solid curves) and numerical~\cite{Blazquez-Salcedo:2016enn} (dots) results. We show the frequencies for axial perturbation (green), polar gravity-led perturbation (blue) and polar scalar-led perturbation (red). The eikonal results are obtained within the small coupling approximation and thus is valid only when $\alpha$ is small.
}
\end{figure}

Let us now compare the above eikonal results with the numerical ones. We focus on EdGB gravity with $f = \exp(\phi) / 4$ and thus $f'_0 = 1/4$. 
Figure~\ref{fig:summary} compares the real QNM frequencies of the eikonal results found here with the numerical ones in~\cite{Blazquez-Salcedo:2016enn} for the three modes as a function of $\alpha$. Notice that the eikonal expressions can accurately describe the numerical results when $\alpha$ is small, where the small coupling approximation is valid. On the other hand, the agreement between eikonal and numerical results for the imaginary QNM frequency is not as good as the real frequency case and requires further study~\cite{Bryant:2021xdh}.

\section{Discussions}

The eikonal method reviewed here can be applied to theories other than GR and sGB gravity. For example, Glampedakis and Silva~\cite{Glampedakis:2019dqh} have applied it to dynamical Chern-Simons gravity~\cite{Alexander:2009tp} which is a parity-violating  gravity. In the action, a parity-violating term that is quadratic in curvature (Pontryagin density) is coupled to a pseudoscalar field. QNM frequencies of BHs in this theory have been computed in~\cite{Yunes:2007ss,Molina:2010fb,Wagle:2021tam,Srivastava:2021imr}. For non-rotating BHs, the background solution is the same as GR (Schwarzschild) and polar perturbation equations are identical to GR. On the other hand, the axial perturbation is coupled to the scalar perturbation, though the axial perturbation potential is identical to the GR Regge-Wheeler potential. Hence, the analysis is much simpler than the sGB case. 

In this chapter, we have focused on non-rotating BHs. A natural extension is to consider rotating BHs. In many theories, analytic BH solutions with arbitrary rotation have not been found yet while slowly-rotating solutions are available. This is indeed the case with sGB gravity~\cite{Pani:2011gy,Ayzenberg:2014aka,Maselli:2015tta} and dynamical Chern-Simons gravity~\cite{Yunes:2009hc,Pani:2011gy,Ali-Haimoud:2011zme,Yagi:2012ya,Maselli:2017kic}. A first attempt on applying the eikonal analysis to slowly-rotating BHs to first order in spin in non-GR theories has been carried out in~\cite{Silva:2019scu}. 

\acknowledgments

We thank Anzhong Wang and Tao Zhu for organizing 2021 Summer School on Early Universe and Gravitational Wave Physics.
We also thank Kostas Glampedakis and Hector Silva for carefully reading the manuscript.
K.Y. acknowledges support from NSF Grant PHY-1806776, NASA Grant 80NSSC20K0523, the Owens Family Foundation, and a Sloan Foundation. 
 K.Y. would like to also acknowledge support by the COST Action GWverse CA16104 and JSPS KAKENHI Grants No. JP17H06358.

\bibliography{master.bib}
\end{document}